\documentstyle[11pt]{article}

\title{{\bf Space-time Autocorrelation and Hubble Flow: \\Probing Small Length Scales in Heavy Ion Collisions }}

\author{{\bf T.A. Trainor and J.G. Reid}\\ \\
{\em Nuclear Physics Laboratory 354290}\\
{\em University of Washington}\\
{\em Seattle, WA 98195}\\
{\em trainor@hausdorf.npl.washington.edu}}
			       
\date{April 18, 2000}

\def\jnl#1#2#3#4{{#1} {\bf #2}, #3 (#4)}

\def\plb{{ Phys. Lett.} B}
\def\prl{ Phys. Rev. Lett.}
\def\prc{{ Phys. Rev.} C}
\def\prd{{ Phys. Rev.} D}

\def\epj{{ Eur Phys. J.} C}

\def\be{\begin{equation}}
\def\ee{\end{equation}}
\def\bea{\begin{eqnarray}}
\def\eea{\end{eqnarray}}

\begin{document}
\maketitle




Two-particle momentum correlations have been applied extensively to the determination of particle source distributions and flow effects in nuclear collisions \cite{wein}. HBT interferometry employs variations of the detected pair intensity in a two-particle momentum space to estimate the conjugate space-time-momentum autocorrelation density of the emitting source. These established techniques depend directly on quantum-mechanical amplitude interference. We propose here an alternative source of momentum correlation induced by a combination of radial or Hubble flow and autocorrelation differences between particle pair types which should provide additional information on the two-particle space-time structure of the emitter. This new method is not primarily an interference effect. In the simplest interpretation it has a cosmological analog in the use of red shifts and a universal Hubble-flow hypothesis to map relative galactic positions.

In this Letter we extend the usual treatment of two-particle momentum correlations to include the possibility of non-chaotic or correlated particle emission from the hadronic freeze-out surface. We adopt a modified two-particle emission density which permits description of nontrivial differences between pair autocorrelations for different pair types. Pair position autocorrelation determines how nuclear flow is sampled by pair partners. If different pair types sample flow in systematically different ways we predict a unique signature in two-particle momentum distributions. 

A general two-particle momentum distribution assuming independent {\em pair} emission can be written as
$ P_2({\bf p}_1,{\bf p}_2) = \int \! d^4x_1d^4x_2 \, S_2(x_1,x_2,p_1,p_2) \left| \Phi_2(x_1,x_2,p_1,p_2) \right|^2 $, where $\left| \Phi_2\right|^2$ may include a distorted-wave treatment to describe Coulomb and final-state interactions (FSI). In what follows we use $x = (x_1 + x_2)/2$ and $y = x_1 - x_2$. If we assume a factorizable Wigner density (chaotic or independent {\em particle} emission), smoothness \cite{urs,uli1} and plane-wave propagation without final-state interactions to a momentum detector we obtain
\begin{eqnarray} \label{pair}
P_2({\bf p}_1,{\bf p}_2) &\approx& P_1({\bf p}_1) P_1({\bf p}_2) + \int \! d^4x_1d^4x_2 \, S_1(x_1,k) \, S_1(x_2,k) \, cos(q \cdot y) 
\end{eqnarray}
which is the conventional elementary expression for the sibling-pair (partners from same event) distribution used in HBT analysis. The second term contains the Fourier transform of the configuration-space {autocorrelation} density. The condition $\int \! d^3p_1/E_1 \, d^3p_2/E_2 \, P_2({\bf p}_1,{\bf p}_2) = N(N-1)$ can be used to normalize this distribution. 

We now seek a more general form of the two-particle Wigner density in which the momentum density is factorized as before but the emission density is not factorized as usual. We assume instead that { pair emission varies locally with partner separation distance} in a way which may depend on some property of the pair, such as total isospin for example. One can think of this as the effect of an initial-state interaction rather than a final-state interaction. Guided by FSI treatments \cite{uli1} it is then natural to factor the two-particle emission density $g(x_1,x_2,p_1,p_2)$ into dependence on mean pair position $x$ and on partner separation $y$, retaining the possibility of parametric dependence on momenta. We then obtain
\begin{eqnarray}
S_2(x_1,x_2,p_1,p_2) &\approx& g_+(x,k) \, g_-(y,q) \, f_1(x_1,p_1) \, f_1(x_2,p_2).
\end{eqnarray}
The argument of
 $ f_1(x_1,p_1) \, f_1(x_2,p_2) = E_1 \, E_2 \, exp\left( - { p_1 \cdot u(x_1) + p_2 \cdot u(x_2)\over T} \right) $ \cite{coop} can be rearranged as
$ p_1 \cdot u(x_1) + p_2 \cdot u(x_2) = k \cdot \Sigma u + {1 \over 2} \, q \cdot \Delta u $, where we adopt the definitions $k = (p_1 + p_2)/2$ , $q = p_1 - p_2$ and define $\Sigma u = u(x_1) + u(x_2)$, $\Delta u = u(x_1) - u(x_2)$. Assuming a longitudinally-invariant radial flow velocity field given near $z = 0$ by $ u(x) = \gamma(x) \, (1,\mbox{\boldmath{$\beta $}}_{\perp}(x)) $ \cite{chap} and particle transverse momentum near $Y = 0$ given by $ p = (m_t,{\bf p}_t) $ we have
$ p \cdot u(x) =  \gamma(x) \, (m_t - \mbox{\boldmath{$\beta $}}_{\perp}(x) \cdot {\bf p}_t ) $. Further assuming that $ \gamma(x) \approx 1$ over most of the volume we can approximate the sum and difference terms as $ k \cdot \Sigma u = m_{t1} + m_{t2} - {\bf k} \cdot (\mbox{\boldmath $\beta$}_{\perp}(x_1) + \mbox{\boldmath $\beta$}_{\perp}(x_2)) $ and
$ q \cdot \Delta u = - {\bf q} \cdot (\mbox{\boldmath $\beta$}_{\perp}(x_1) - \mbox{\boldmath $\beta$}_{\perp}(x_2)) $. 

For discussion purposes we represent $P_2({\bf p}_1,{\bf p}_2)$ as a sum of two terms $ A({\bf k,q}) + B({\bf k,q}) $, where the second (interference) term contains the usual HBT Fourier transform ($= 0$ in the case of mixed pairs). The first term, which we now examine in detail, is ordinarily written as the product of two single-particle distributions. We now approximate radial flow observed in HI collisions \cite{na49} by analogy with cosmological Hubble flow and assume that\mbox{ \boldmath $\nabla \beta$} $ = H$\mbox{ \boldmath $\cal I$} 
from which follows
\mbox{$\Delta$ \boldmath $\beta$}$(x) \approx H \, \Delta {\bf x}$, that is, relative transverse velocity is proportional to transverse separation for any two points in the emitting system. This approximation may be unrealistic at the scale of a nuclear diameter but quite reasonable at the hadron scale of primary interest here. Combining results we have for $A({\bf k,q})$ the product of three factors
\begin{eqnarray} \label{threefac}
A({\bf k,q})  &=& m_{t1} \, m_{t2} \, e^{-{m_{t1} + m_{t2} \over T}} \cdot \int \! d^4x g_+(x,k) \, exp\left( {2 H \over T }\, {{\bf k} \cdot {\bf x} } 
\right) \cdot \int \! d^4y \,   g_-(y,q) \, exp\left({{H \over 2 T}  \, {{\bf q} \cdot {\bf y} } } \right)
\end{eqnarray}
The first two factors are equivalent to the single-particle radial-flow case
for the particle pair as a unit. The third factor contains potentially new information about the pair autocorrelation density $g_-(y,q)$.

The two-particle correlator \cite{chap,na491} can be expressed  as $ C_2({\bf k,q}) = a({\bf k,q}) + b({\bf k,q}) $, where for instance
$ a({\bf k,q}) = A_{obj}({\bf k,q}) / A_{ref}({\bf k,q}) $ and $b({\bf k,q})$ contains the HBT Fourier transform from which the large-scale source geometry is inferred. In the ratio $a({\bf k,q})$ the first two factors of Eq. (\ref{threefac}) cancel, leaving the ratio of two integrals on particle separation. The standard assumption of chaotic particle emission implies that the autocorrelation densities for object and reference systems are equal ($g_{obj}(y,q) \approx g_{ref}(y,q)$) and therefore $a({\bf k,q}) \approx 1$. We argue that this assumption is arbitrary and that the consequences of $g_{obj} \neq g_{ref}$ should be tested experimentally. For the ratio we then obtain (we have changed notation: $g_- \rightarrow g_{\stackrel{obj}{ref} } $)
\begin{eqnarray} \label{aa}
a({\bf k,q})  &\approx& 1 +  {1 \over  V} \, \int \! d^4y \,   \{ g_{obj}(y,q) - g_{ref}(y,q) \} \, cosh \left( {H \over 2 T} {\bf q} \cdot {\bf y}   \right) 
\end{eqnarray}
where $V$ is the volume inferred from the autocorrelation distribution, and we argue that the $sinh$ term in the integrand is suppressed by symmetry, leaving $cosh$ as the dominant term.

Put simply, just as radial or Hubble flow causes broadening (blue shift) of the single-particle ${\bf p}_t$ distribution and  two-particle ${\bf k}$ distribution to higher momenta it also distorts the distribution on pair momentum { difference} ${\bf q}$. However, in the latter case the degree of distortion depends on the autocorrelation density on partner separation as opposed to mean position on the emitter. If this autocorrelation is different for object and reference distributions (Hubble flow is sampled differently in the two cases) the difference is reflected to lowest order in a large-scale {\em quadratic} departure from unity in the ratio of the two distributions on ${\bf q}$. Again, this is not primarily an interference effect.

We can model the difference $ g_{obj}(y,q) - g_{ref}(y,q)$  by assuming that event-mixed (reference) pairs follow an autocorrelation density on the transverse emitting space described by a gaussian with $rms$ transverse radius $R$ reflecting the size of the emitting system. We further assume that for sibling pairs the autocorrelation density may be modified by a second smaller concentric gaussian with radius $r  < R$, added or subtracted (correlation or anticorrelation) depending on pair properties, and with amplitude $\epsilon \ll 1$ reflecting the strength of this pair (anti)correlation. Normalization of densities is maintained.   Source radii inferred from HBT analysis, sometimes referred to as `lengths of homogeneity' \cite{siny}, more generally correspond to the second moments (covariance matrix) of the autocorrelation density which may have an arbitrary shape near the origin, thus precluding use of the curvature matrix \cite{curv} to represent source sizes.
 
The leading term after expanding the $cosh$ in Eq. (\ref{aa}) contains the factor $<\!({\bf q} \cdot {\bf y})^2\!>_{\Delta g}$. We can express the dot product in terms of longitudinal and transverse components for ${\bf y}$ and ${\bf q}$ referred to mean vectors ${\bf x}$ and ${\bf k}$ respectively. The cross term linear in $q_l\,y_l$ is zero by symmetry. This leaves the two terms $<(q_t\, y_t)^2>$ and $<(q_l\, y_l)^2>$. We assume that the autocorrelation difference $\Delta g(y,q)$ localizes $y_t$ and $y_l$ to a common $rms$ value $\hat y \equiv r$, and that $\Delta g(y,q)$ may contain a strong correlation between $\phi_y$ and $\phi_q$, the opening angles between position and momentum vector pairs respectively, in which case we obtain $<({\bf q}\cdot {\bf y})^2>_{\Delta g} \approx r^2\{ 4 k^2 <sin^2(\phi/2)> + q^2 <cos^2(\phi/2)> \}$, where $k$ and $q$ are now the scalar mean and difference of $p_{t1}$ and $p_{t2}$, and $\phi$ is the azimuthal opening angle between momentum vectors. Gathering results we obtain
\begin{eqnarray} \label{akq}
a({p_{t1},p_{t2}})  &\approx& 1 \mp   \epsilon \, { \hat \beta^2_{\perp} \over 8 T^2} \, (r/R)^2 \, \{1 - (r/R)^2\} \,  \{2 k^2\, (1 - <\!cos(\phi)\!>_{\Delta g})  \nonumber \\
& & + ~q^2 \, (1 + <\!cos(\phi)\!>_{\Delta g})/2 \} ~~~~~~<\!cos(\phi)\!>_{\Delta g} \in [0,1]
\end{eqnarray}
where $\hat \beta_{\perp}$ is the $rms$ flow speed for the $g_{ref}(y,q)$ distribution, and $<\!cos(\phi)\!>_{\Delta g}$ represents the autocorrelation-averaged pair momentum opening angle. The negative resp. positive sign corresponds to correlation resp. anticorrelation of sibling pairs on separation distance relative to a mixed reference. Thus, any difference in the autocorrelation density of emitted particles for sibling pairs relative to a mixed reference (breakdown of chaotic-source assumption) as measured by a difference in $rms$ partner separation between sibling and mixed pairs combined with local Hubble flow should result in a contribution to the correlator which is quadratic in $p_t$ mean and/or difference. Eq. (\ref{akq}) should model any emission scenario from uncorrelated emission throughout a transparent volume to normal particle emission from an opaque cylinder.

The value of $<\!cos(\phi)\!>_{\Delta g}$ could be established from experiment by comparing the amplitudes of observed quadratic dependences on $k$ and $q$. Due to the finite mean free path of pions in nuclear matter and possible opacity effects \cite{heis} the mean value of the cosine may be significantly different from zero, and possibly close to unity. For example, the relationship between partner separation and opening angle for normal emission from an opaque cylinder of radius $R$ is $cos(\phi)(y) = cos\{ 2 cos^{-1}\sqrt{1 - (y/2R)^2}\}$. This expression remains close to unity ($> 0.9$) even for substantial separation ($y/R < 0.5$). The implications of opacity for source space-time structure is a very active area of research. Information on opacity has been derived by comparing $R_{out}$ and $R_{side}$ obtained from HBT analysis. However, such comparisons are ambiguous due to the well-known cross talk between source duration and thickness in the `out' direction. Determination of $<\!cos(\phi)\!>_{\Delta g}$ by this new method could provide important independent information on the issue of opacity. Currently there is unresolved conflict between models describing a `sudden transition to full transparency' \cite{uli1} ($<\!cos(\phi)\!>_{\Delta g} \approx 0$) and models describing sustained emission from a thin surface layer on an opaque source \cite{heis} ($<\!cos(\phi)\!>_{\Delta g} \approx 1$). Alternatively, the value of $<\!cos(\phi)\!>_{\Delta g}$ may be influenced by resonance decay kinematics.

In summary, the large-$q$ region of two-particle momentum space may carry significant dynamical information, may in fact be oversubscribed by different dynamical phenomena at some level of
sensitivity. Analysis of the two-particle momentum correlator in this region could provide detailed information on the emitting surface beyond the large-scale features of the space-time-momentum density revealed by established HBT interference techniques. We here propose a new strategy  combining nuclear Hubble flow with a generalized emitter autocorrelation density to
probe details of the hadronic freeze-out surface at small length
scales, especially to explore the dependence of the emission
process on particle-pair characteristics such as total isospin. We find
that any departure from the chaotic emission assumption used in
standard HBT analyses combined with local Hubble flow should result in a quadratic deviation from unity of the two-particle correlator at large transverse momentum sum and/or
difference. By observing such correlations one may learn about
two-particle spatial and angular correlations at small length scale on the
emitter which could indicate whether hadrons are generated from a bulk
prehadronic state (QGP) or decouple with correlations typical of a hadronic
cascade.

TAT thanks U. Heinz for reading a draft of this paper and for helpful discussions.
Some of this work was initiated with the support of RHIC
R\&D grant DE-FG06-90ER40537.  We are thankful for continuing research
support from the USDOE under grant DE-FG03-97ER41020.

\end{document}